# New Power Decoupling Method for Grid Forming Inverter Based on Adaptive Virtual-Synchronous Machine in Weak Grids

Waleed Breesam and Stefan M. Goetz

*Abstract*—Many countries' policies have shifted rapidly towards using renewable energy for climate reasons. As a result, inverter-based resources are beginning to dominate power systems. Key elements for managing the loss of conventional generators are virtual-synchronous-generator-based grid-forming inverters. Despite the unique advantages of this technology, there are still various challenges, most notably the problem of active–reactive power coupling due to a nonzero power angle and high grid impedance ratio *R/X*. The effect of power coupling means that any change in the inverter's active power will affect the reactive power and vice versa. This challenge results in grid instability, reduces control performance, and restricts the active power delivery capability of the inverter to the grid. This paper presents a new vision to solve this impact in weak grids by a new power-decoupling method based on adaptive virtual synchronous generator parameters. The power coupling will be studied considering the parameters causing this effect. Fuzzy logic will serve to adjust the parameters of power control loops. Hardware-in-the-loop testing on a real-time simulator (OP4610) and a physical microcontroller verified and validated the proposed method. The results showed the proposed method's effectiveness in eliminating static and dynamic power coupling and improving the grid-forming inverter performance under different operating conditions.

*Index Terms*—Adaptive virtual synchronous generator, Fuzzy logic control, Hardware in the loop, grid forming inverter, grid impedance, grid impedance ratio, power coupling.

## I. Introduction

Inverter-based resources (IBRs), such as wind and photovoltaics power, play a significant role in responding to environmental problems and energy crises. Currently, the popular operation of IBRs is grid-following inverters. Alternatively, grid-forming inverters are becoming gradually popular due to their higher performance and ability to operate in standalone and grid-connected modes. Nevertheless, the IBRs typically connect with the power grid, which has a weak frequency and voltage support capability due to a lack of inertia and damping characteristics.

The virtual synchronous generator (VSG) concept is proposed to address these challenges. VSG mimics the dynamic behaviour of a synchronous generator through a swing equation to increase the inertia and damping of IBRs and improve the transient response characteristics [1, 2].

Despite this feature, VSG still suffers from challenges, one of which is power coupling, in which any change in the output active power will affect the output reactive power and vice versa. The reason is due to the impact of nonzero power angle and high grid impedance ratio *R/X*. The power coupling exists in a low-voltage grid because of the high *R/X* ratio of grid impedance. The decoupling between the active and reactive power is at present still mostly satisfied in medium or high-voltage grids where the grid impedance ratio *R/X* is small enough. The power coupling can lead the power system to stability degradation (transient oscillations) and produce a steady-state error in reactive power, thus increasing the reactive power absorption to provide active power support [3].

Power decoupling approaches can be classified based on their working principle into four main categories: coordinate transformation methods [4, 5], virtual impedance methods [6, 7, 8], feedforward-compensation methods [9, 10, 11, 12, 13, 14], and hybrid methods [15, 16, 17].

Coordinate-transformation-based methods use the grid impedance angle to rotate the active and reactive power vectors to form a virtual power decoupling, which can then be controlled by voltage and frequency. The same technique is used for the virtual voltage frequency. The above methods were applied based on droop control, which is considered simple and does not contribute much to the power coupling. At the same time, they are not suitable for VSG control since these control systems contribute mainly to increasing the power coupling due to oscillation characteristics [18].

Virtual-impedance-based methods, on the other hand, are widely used due to their ease of implementation and clear physical meaning. This method works by adding a virtual impedance (usually inductive) at the inverter output through a voltage control loop to make the grid impedance characteristics more inductive, thus reducing power coupling. However, the above methods suffer from the limitation of power decoupling even when the optimum values of the virtual inductance are determined. Moreover, they did not consider the effect of varying grid impedance and *R/X* ratio. In addition, the effectiveness of the methods used to reduce the power angle is unclear.

The above two approaches are indirect because they do not provide a direct decoupling mechanism through the power control loops. Conversely, feedforward compensation-based methods add an action that eliminates or compensates the effect of the power coupling in the active and reactive power control loops. Therefore, the methods used in this approach can be considered direct.

One technique adopted in this approach is using a diagonal decoupling matrix in the power loop to eliminate or compensate the component responsible for power coupling [9]. However, this method is a linearized power model-based method. Thus, it does not reflect or represent the actual behaviour of the power loop and does not consider the variation in grid impedance and R/X ratio.

On the other hand, Wang et al. propose a novel power decoupling technique based on total sliding mode control to generate voltage compensation in the reactive power loop [10]. Despite its effectiveness, there are still some drawbacks. First, it relies on the fixed value of the grid impedance, which is not accurate in actual situations. Second, an online-trained adaptive fuzzy neural network was implemented to mimic the sliding mode control law to relax the requirement for detailed system information. This, in turn, will add complexity to the design of the proposed method, which can be easily avoided without the need for such complications, as the idea of this paper will demonstrate later.

Another interesting method in this approach is using an extended state observer, which considers the power coupling calculation (virtual power angle and virtual voltage) as internal disturbances to compensate for the coupling effect [11]. However, this method adopts an analytical and mathematical approach to estimate these internal disturbances within a specific operation range for compensation. More precisely, the method requires building a small signal model of the power coupling behaviour (the model is linearizing around an operating point). Thus, these assumptions may not reflect the actual considerations over a wide operating range, especially in sudden and significant disturbances. Moreover, the variation in *R/X* has not been adopted or clarified. In the same context, a discrete-time multivariable controller based on the prevalent control structures for grid-forming inverters is proposed, and an H-infinity method is used to tune the parameters of the proposed controller [12]. The drawback of this method is the optimization problem. This method requires tuning the controller parameters, necessitating using an H-infinity method to address this problem, thus adding complexity to the design process based primarily on mathematical analysis and prior knowledge of the system. Furthermore, the effectiveness of the proposed control was tested at a fixed grid impedance value and not tested under different values.

The other methods mentioned in this approach applied full-state feedback and inverse coupling characteristic matrix [13, 14]. However, these methods were applied based on droop control, which does not contribute much to the power coupling compared to VSG control. Therefore, as we mentioned before, it is not appropriate and cannot be adopted in VSG control.

Some literature combines methods from the above approaches to increase the effectiveness of power decoupling, and thus, they can be considered hybrid methods. Liu et al. use a power feedforward link and modify the virtual damping coefficient for coupling compensation and power oscillation suppression [16]. Again, the design of the proposed method is based on linearized power coupling behaviour, and the changes in the grid parameters are not considered. On the other hand, a unified calculation method by the overlapped eigenvalues distribution is proposed to identify the equivalent of inertia and damping. Also, virtual inductance control can enhance and improve both inertia and damping based on this equivalence model [17]. In addition to the proposed method's reliance on prior mathematical knowledge, another drawback is that it adopts a fixed value for the grid impedance. Furthermore, the values of the *R/X* ratio are chosen within the range of a grid impedance considered inductive and not within a grid with a resistive effect.

Another perspective can provide a solution that adopts the impact of the VSG parameters on the power coupling. Li et al. report that changing the virtual parameters of power control loops significantly affects the strength of power coupling [18]. They concluded that the droop parameter in the reactive power control loop directly affects the static coupling (steady-state error in the reactive power). In contrast, the damping and inertia parameters in the active power control loop notably impact the dynamic coupling (during the transient and steady-state process). Hence, the importance of adjusting these virtual parameters and their role in power decoupling is emphasized.

Since VSG-based grid-forming inverters are nonlinear and influenced by several parameters and disturbances, there is a need for a power-decoupling method or technique that does not depend primarily on analytical assumptions or prior knowledge of the mathematical model of the system. Smart techniques such as fuzzy logic control could achieve power decoupling. This technique can function with complex and nonlinear systems that require prediction and adaptation under different operating conditions. In addition, they do not rely on modelling and thus do not require prior knowledge of the controlled system (model-free-based techniques) [10]. Motivated by this, this paper proposes a new method to adapt power control loop parameters to provide power decoupling capability in VSG-based grid-forming inverters. Therefore, the classification of power decoupling approaches can be updated, as shown in Fig. 1.

The main contribution of this paper is as follows:
1- A new method called adaptive power control loop parameters is proposed. This method uses the fuzzy logic controller technique to dynamically adjust and adapt the control parameter in the reactive power control loop to eliminate the steady state error in reactive power (static coupling) and, thus, provide the inverter's active power delivery capability to the grid.

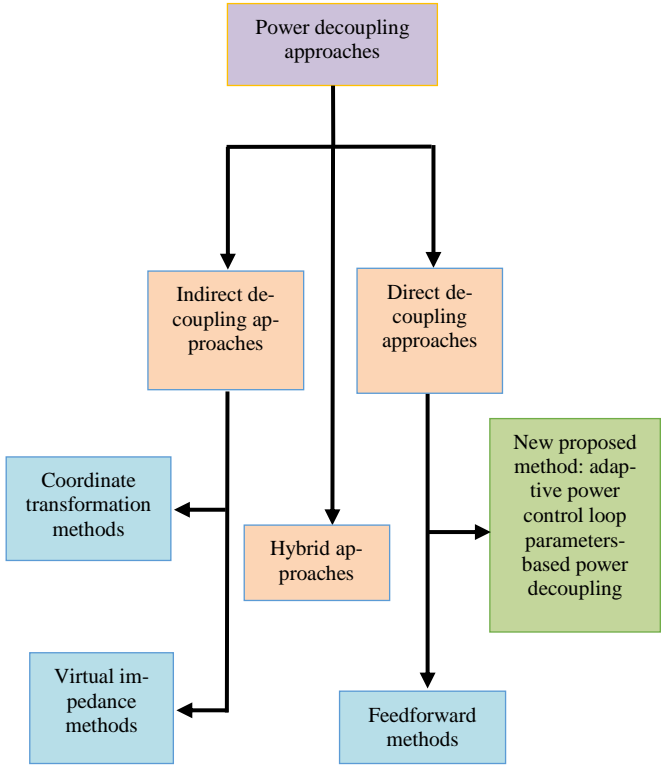

Fig. 1. Classification of power decoupling approaches.

2- The fuzzy logic controller removes the dynamic coupling by adapting the active power control loop's virtual parameters (inertia and damping). The adaptation allows the grid-forming inverter to eliminate fluctuations and ensures a smooth, active power response during transients and disturbances to stabilize the grid.

3- The proposed method represents a new generation of model-free virtual control. This method does not require prior knowledge of the system model nor requires further calculations or mathematical analytics, indicating its simplicity (less complexity) and giving it an additional advantage in the design stage. We experimentally tested and validated the control algorithm using hardware-in-the-loop (HIL) based on the real-time simulator (OP4610) and microcontroller.

The rest of this paper is structured as follows: Section II presents the traditional VSG-based grid-forming inverter model. Section III studies and analyzes the characteristics and issues of power coupling. Section IV describes the proposed power decoupling method using a fuzzy logic controller based on adaptive virtual parameters. Section V presents the results and discussions. Finally, Section VI provides the conclusions of the paper.

## II. PRINCIPLE OF VSG-BASED GRID FORMING INVERTER

Fig. 2 illustrates a typical block diagram of a VSG-based grid-forming inverter. This three-phase inverter is connected to the grid through an LC filter. $(i_a, i_b, i_c)$ and $(v_a, v_b, v_c)$ are the output voltages and currents of the inverter at the point of the common coupling (PCC), while $(i_{al}, i_{bl}, i_{cl})$ represent the inverter currents passing through the filter inductance. The equivalent impedance of the grid can be expressed as $(Z_g = R_g + jX_g)$, where $R_g$ is the grid resistance, and $X_g$ is the grid inductive reactance. Fig. 2 also includes the control units, which consist of several parts and loops. The power control loops are fed with the measured active and reactive power values. The voltage and current control loops regulate the voltage and current of the inverter by tracking the reference frequency and voltage values generated by the power control loops, providing protection to the physical components of the inverter and ensuring its operation is within the permissible range. The power control loops are a primary control where VSG is used to virtually mimic the characteristics of a synchronous generator through the swing equation. Fig. 3 presents the block diagram of the power control loops.

The power control loops follow [10, 19]

$$J\omega_o \frac{d\omega}{dt} = P_m - P_e - D(\omega - \omega_g), \quad (1)$$

$$E = E_o + k_q(Q_m - Q_e), \quad (2)$$

where $P_m$ and $P_e$ respectively are the rated active power and the output active power of the inverter; $J$ and $D$ respectively denote the virtual moment of inertia and virtual damping coefficient; $\omega_g$ is the rated angular frequency of the grid; $\omega_0$ is the nominal angular frequency; and $\omega$ is the angular frequency reference generated by the VSG. $E_r$ represents the voltage reference generated by the VSG, $E_o$ is the rated voltage, $Q_m$ and $Q_e$ are the rated reactive power and the output reactive power of the inverter, respectively, and $k_q$ is the droop coefficient of the reactive power control loop.

## III. CHARACTERISTICS AND ISSUES OF POWER COUPLING

The grid-forming inverter can operate in islanding (autonomous) and grid-connected mode. As we mentioned earlier, the grid-forming inverter in the second mode is influenced by the characteristics and parameters of the grid. More precisely, the strength of grid impedance will significantly affect the system's stability. The situation worsens when using VSG-based control due to this approach's dynamic characteristics, which will cause more challenges. Fig. 4 presents an equivalent circuit of the system.

The power flow follows from the inverter to the grid follows [10, 12]

$$P_e = \frac{3}{Z}[E_r^2 \cos\alpha - V_g E_r^2 \cos(\alpha + \theta_r)], \quad (3)$$

$$Q_e = \frac{3}{Z}[E_r^2 \sin\alpha - V_g E_r^2 \sin(\alpha + \theta_r)], \quad (4)$$

$$Z = \sqrt{R_g^2 + X_g^2}, \quad (5)$$

$$\alpha = \tan^{-1}\left(\frac{X_g}{R_g}\right), \quad (6)$$

where $Z$ and $\alpha$ are the magnitude and the angle of grid impedance, respectively; $V_g$ is the amplitude of the grid voltage; $\theta_r$ is the phase angle of the PCC voltage or power angle.

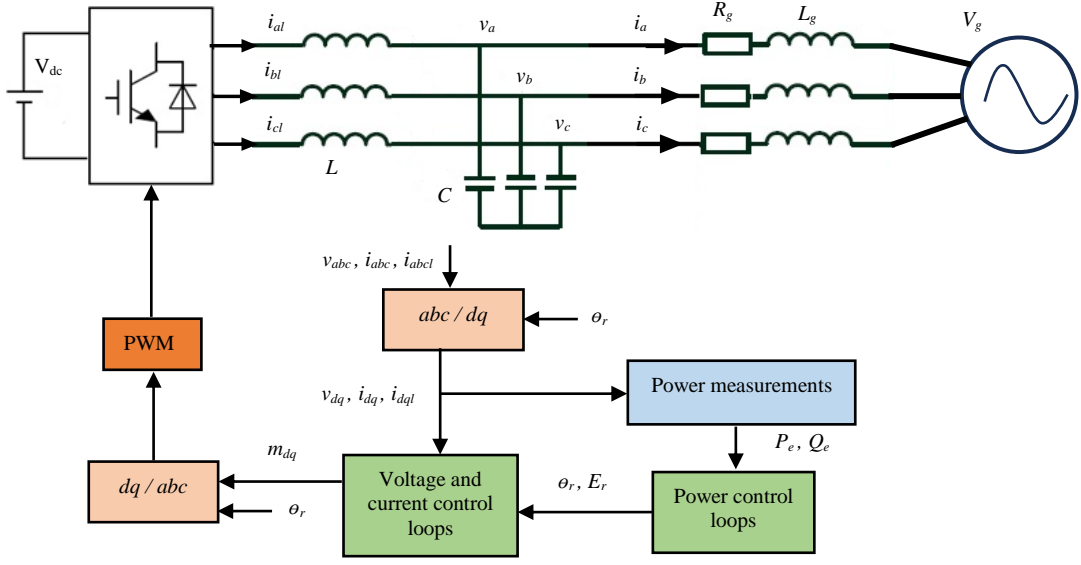

Fig. 2. Typical block diagram of a VSG-based grid-forming inverter.

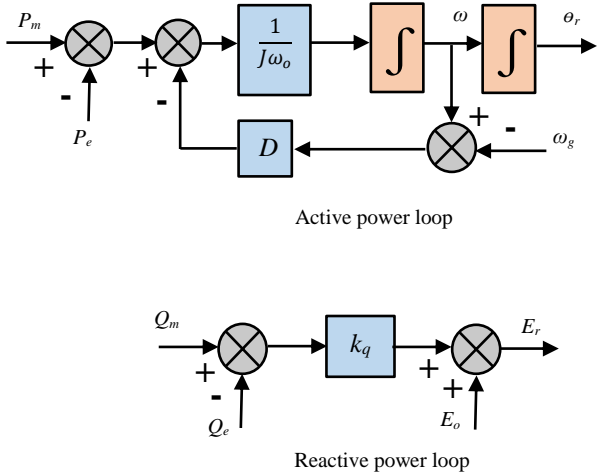

Fig. 3. Block diagram of the power control loops.

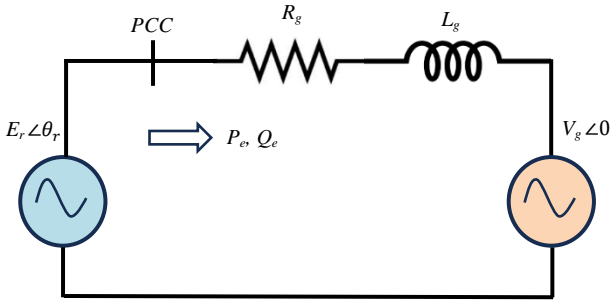

Fig. 4. Equivalent circuit of the system.

the partial differential of (3) and (4) with respect to $E_r$ and $\theta_r$ and linearition around an operating point ($E_{ro}$, $\theta_{ro}$) results in a small signal model can be derived as

$$\begin{bmatrix} \Delta P_e \\ \Delta Q_e \end{bmatrix} = \begin{bmatrix} G_1 & G_2 \\ G_3 & G_4 \end{bmatrix} \begin{bmatrix} \Delta \theta r \\ \Delta E r \end{bmatrix}, \quad (7)$$

where the parametric models for $G_1$, $G_2$, $G_3$, and $G_4$ are obtained as

$$G_1 = \tfrac{3}{Z} [V_g E_{ro} \sin(\alpha + \theta_{ro})], \quad (8)$$

$$G_2 = \tfrac{3}{Z} [2E_{ro} \cos\alpha - V_g \cos(\alpha + \theta_{ro})], \quad (9)$$

$$G_3 = \tfrac{-3}{Z} [V_g E_{ro} \cos(\alpha + \theta_{ro})], \quad (10)$$

$$G_4 = \tfrac{3}{Z} [2E_{ro} \sin\alpha - V_g \sin(\alpha + \theta_{ro})], \quad (11)$$

Notably, the above equations show that controlling the active and reactive power is not possible independently due to the impact of grid parameters ($Z$ and $\alpha$) in addition to $\theta_{ro}$, which leads to their interaction. In other words, any change in the output active power from the inverter will affect the reactive power and vice versa. Fig. 5 illustrates more clearly the impact of power coupling.

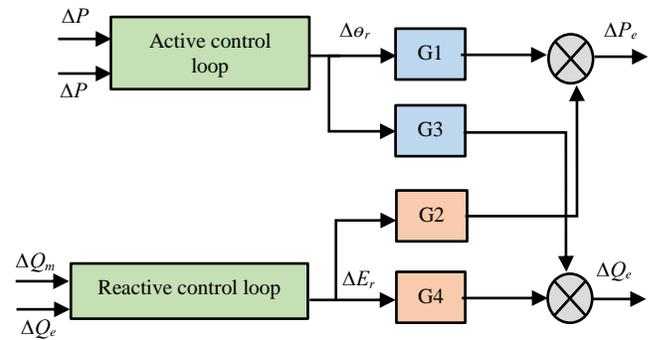

Fig. 5. Small-signal model of power control.

If $X_g \gg R_g$, this leads to no power coupling. Conversely, if $R_g \gg X_g$, or $R_g \approx X_g$, this means that there is a power coupling that increases as the resistance value increases. Indeed, even if the inductive grid impedance characteristics are achieved, the power angle will still be affected if it is large enough and cannot be neglected. More precisely, the power angle will increase as the line inductance and output power from the inverter increase, aggravating the power coupling [10][12].

The grid impedance value can also indicate the grid strength or short circuit ratio (SCR), which in turn indicates the inverter's operating range. The SCR can be defined as the inverse of grid impedance per unit [20][21][22]

$$\text{SCR} = \frac{1}{Z\,(p.u.)} = \frac{V_g^2}{Z\,P_{inv,rated}}, \quad (12)$$

where $P_{inv,rated}$ is the rated power of the inverter.

Thus, the characteristics and strength of the grid are directly proportional to the system voltage level and inversely proportional to both the grid impedance and the rated power of the inverter. The grid is considered weak if 2<SCR<3, and strong if SCR>3. The change in $R/X$ ratio, as well as varying the grid impedance value, will deteriorate the dynamic performance of the VSG-based inverter and yield possible instability risk [3][23].

## IV. THE PROPOSED POWER DECOUPLING METHOD

Based on the power coupling effect analysis in the previous section, a strategy must be provided to eliminate or minimize this effect in the VSG-based inverter. Besides, several studies indicate that despite the control approach, using VSG may aggravate the power coupling and cause instability; at the same time, if the VSG parameters are selected appropriately, it will give a robust solution in power decoupling and enhance the control performance [3, 18]. Meanwhile, the proposed method must meet another specific criterion for outperforming other methods. For example, the proposed method necessity deals with scenarios of changing the grid parameters and reducing the power angle. It does not require prior knowledge of the system model (model-free). To this end, a new method for power decoupling is adopted based on adjusting the VSG parameters dynamically, i.e., the $k_q$ parameter in the reactive power control loop, $J$ and $D$ in the active power control loop. A fuzzy logic technique adapts these virtual parameters; thus, the proposed method's classification is within a direct decoupling approach.

The idea of fuzzy systems presents influential control techniques for dealing with nonlinear systems by using input-output data from the original mathematical description of the system. Fuzzy rules, which are determined based on the designer's information and knowledge, are if-then fuzzy sets, inference, and logic. These rules play a vital role in representing expert control knowledge in linking the input variables of the fuzzy logic controller to the output variables [24]. The actual or crisp values of the inputs are represented as a linguistic variable within specific ranges, and this process is called fuzzification. Many curve patterns (membership functions) can be used to fuzzify crisp input values, such as triangles and Gaussians. This procedure is followed by executing the rules by performing a set of logical operations and then converting the result into crisp values through defuzzification. The inputs of the fuzzy logic unit are divided into fuzzy subsets that take linguistic variable titles [25, 26].

It is important to note that adjusting $k_q$ alone is sufficient to reduce static power coupling and eliminate the steady state error in reactive power. However, if the $R/X$ increases during the adaption of the $k_q$ parameter, an oscillation (overshoot) in the active power response will begin to appear and thus needs adapting the parameters of the active power loop to improve the power response and maintain high control performance of the inverter under different operating conditions. Fig. 6 shows the structure of the proposed method. The inputs for the fuzzy logic controller are $Q_e$ and $R/X$, while $k_q$, $D$, and $J$ are the outputs.

Tables I, II, and II show the fuzzy rules used to achieve power decoupling in the system by adapting $k_q$, $D$, and $J$ parameters. where N is negative, VN is very negative, Z is zero, L is large, VL is very large, S is small, M is medium, and H is high.

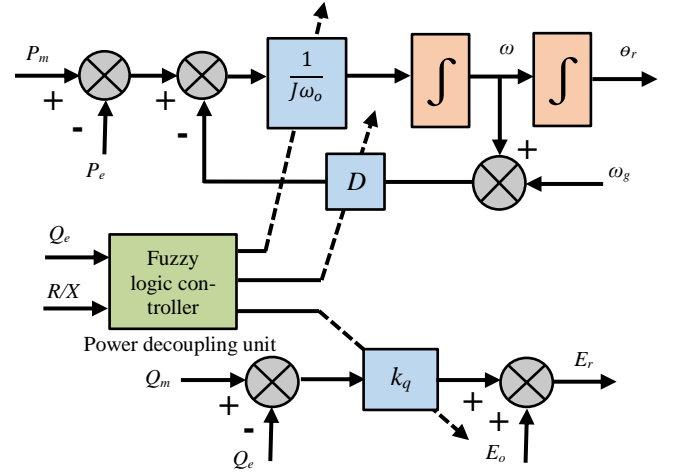

Fig. 6. Structure of the proposed method.

Table I
Fuzzy rules of virtual $J$ control for linguistic variables

|  |  | $Q_e$ | | |
|---|---|---|---|---|
|  |  | VN | N | Z |
| R/X | M | S | M | L |
|  | H | S | S | M |

Table II
Fuzzy rules of virtual $D$ control for linguistic variables

|  |  | $Q_e$ | | |
|---|---|---|---|---|
|  |  | VN | N | Z |
| R/X | M | VL | L | M |
|  | H | VL | VL | L |

Table III
Fuzzy rules of virtual $k_q$ control for linguistic variables

|  |  | $Q_e$ | | |
|---|---|---|---|---|
|  |  | VN | N | Z |
| R/X | M | L | M | S |
|  | H | L | L | M |

## V. Results and Discussion

We implemented a hardware-in-the-loop (HIL) setup is used based on a real-time simulator (OPAL RT OP4610) to verify and validate the proposed method (Fig. 7). The system model (VSG-based grid forming inverter) is implemented using MATLAB Simulink on the host PC and loaded into the target according to the parameters listed in Table IV. It is run in real-time with a step size of 10 μs. The system operates in the following ranges: $1 \leq R/X \leq 2.5$ and $0.85 \leq Z_g \leq 2.6$, which is equivalent to $0.96 \leq SCR \leq 2.87$. The control algorithm, including the power decoupling unit, runs in a microcontroller based on an ARM system (SAM3X8E, Cortex-M3). The host PC monitors and records the target and microcontroller's data, waveform, and responses.

Table IV
Parameters of the nominal system

| Symbol | Description | value |
|---|---|---|
| $V_{dc}$ | Inverter dc-link voltage | 700 V |
| $V_g$ | Grid voltage (phase) | 220 V |
| $L$ | Inverter filter inductor | 1.6 mH |
| $C$ | Inverter filter capacitor | 1500 μF |
| $L_g$ | Grid inductance | 4 mH |
| $R_g$ | Grid resistance | 1.25 Ω |
| $J$ | Virtual inertia constant | 50 W.s$^2$/rad |
| $D$ | Damping coefficient | 90000 W.s/rad |
| $k_q$ | Droop coefficient of reactive power loop | 0.005 V/Var |
| $\omega_o$ | Angular frequency | 314 rad/s |
| $f$ | Switching frequency | 6 kHz |
| $P_m$ | Rated active power | 20 kw |
| $Q_m$ | Rated reactive power | 0 Var |

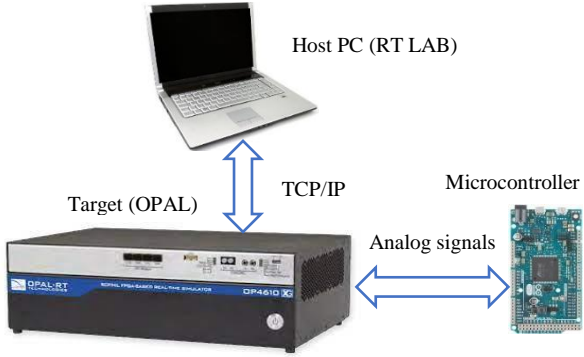

Fig. 7. The setup and configuration of the HIL platform.

Fig. 8 shows the Gaussian membership functions for the fuzzy logic controller's inputs. In this paper, we adopted the centre average method for the defuzzification, where the fuzzy logic controller's output ranges respectively are $J = [20, 45]$, $D = [140000, 190000]$, and $k_q = [0.1, 0.5]$.

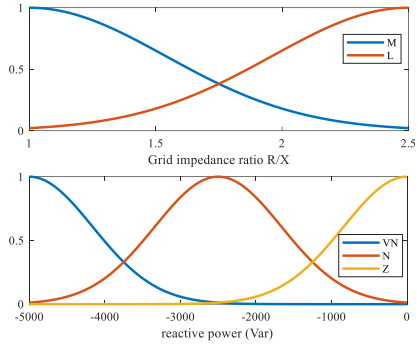

Fig. 8. Membership functions of fuzzy logic control inputs.

### A. Without power decoupling

The VSG-based grid-forming inverter operates without the power decoupling method at $L_g=0.95$mH and $R_g=0.75$ Ω ($R/X=2.5$). Fig. 9 shows the power responses of the inverter at PCC and the grid. The results show a significant power coupling, as any change in the active power is accompanied by a change in the reactive power. Moreover, there is a significant steady-state error in reactive power response. As a result, this causes the inverter to absorb more reactive power and thus degrades the grid.

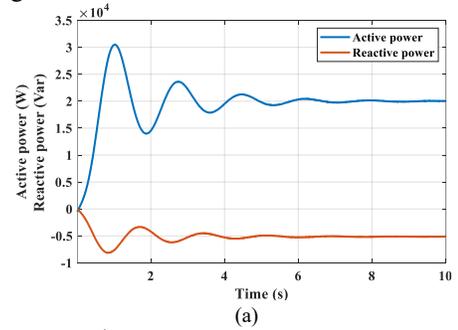

(a)

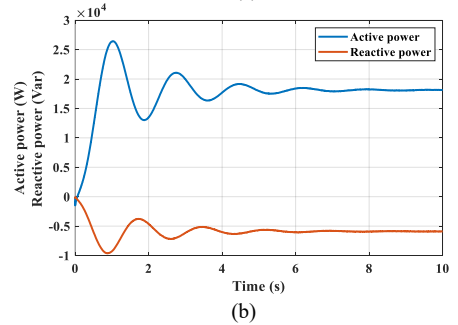

(b)

Fig. 9. Power responses without power decoupling: (a) at PCC (b) at grid.

## B. With the proposed power decoupling method

### 1- Case 1: *R/X*=1

Fig. 10 shows the power responses of the inverter using the proposed method at nominal grid parameters, i.e., $L_g$=4mH and $R_g$=1.25 Ω. The adjustment of $k_q$ will affect the active power response (Fig.10(a)). Therefore, adjusting $J$ and $D$ improves the active power response, especially when *R/X* values increase (Fig.10(b)). This gives the grid-forming inverter the benefit of eliminating fluctuations and ensuring smooth, active power response during transients and disturbances, thus contributing to the grid's stability under varying operating conditions. Fig. 11 and 12 show the power responses when grid impedance parameters deviate from the nominal value. It is clear that the proposed method successfully eliminates the coupling impact and improves the responses when the grid impedance values change.

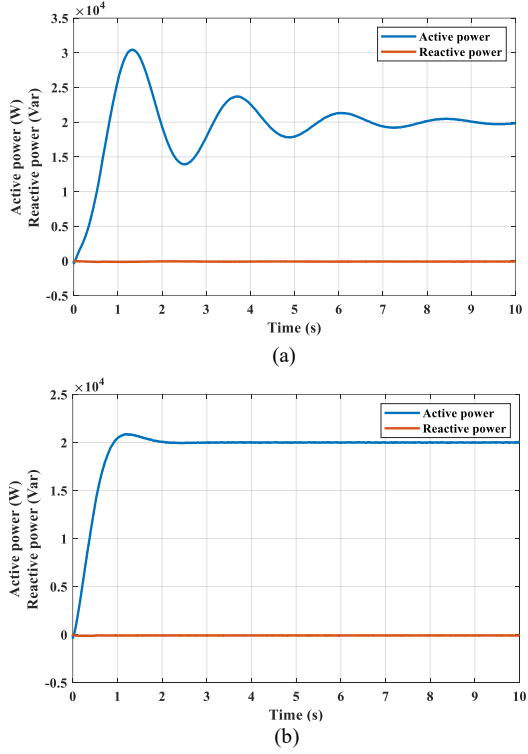

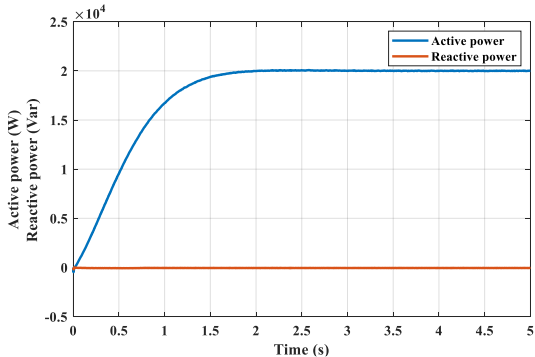

Fig. 10. Power responses with power decoupling: (a) without adjusting *J* and *D* (b) with adjusting *J* and *D*.

Fig. 11. Power responses at PCC: $L_g$=2.3mH and $R_g$=0.75 Ω.

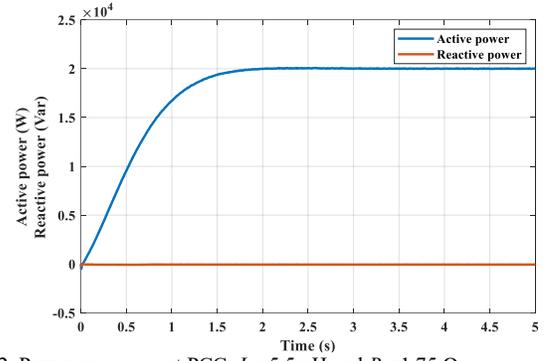

Fig. 12. Power responses at PCC: $L_g$=5.5mH and $R_g$=1.75 Ω.

### 2- Case 2: R/X=2.5

In this case, the proposed method's effectiveness will be tested when the *R/X* ratio increases (the grid impedance characteristics are more resistant, and thus, the coupling becomes worse). Power responses at different grid impedance values are shown in Fig. 13 and 14, respectively. Again, the results demonstrate the robustness of the proposed method in eliminating the effect of power coupling under different operating conditions.

On the other hand, Fig. 15 shows the power angle with and without using the proposed power decoupling method. The results show that the proposed method also reduced the power angle because, as mentioned before, a high-power angle aggravates the coupling effect. Table V shows the power angle values for the other cases under test.

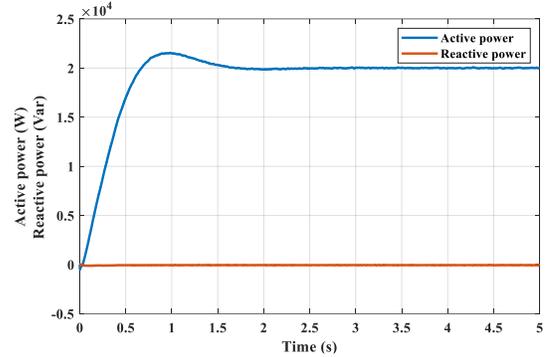

Fig. 13. Power responses at PCC: $L_g$=9.5mH and $R_g$=0.75 Ω.

In the same context, table VI summarizes the percentage increase in the amount of power transferred from the inverter to the grid for all the above cases compared to the presence of power coupling. The proposed method increased the inverter's ability to deliver the maximum possible power to the grid, especially when the grid impedance and *R/X* increase. For convenience, Table VII summarizes the comparison among various power decoupling methods.

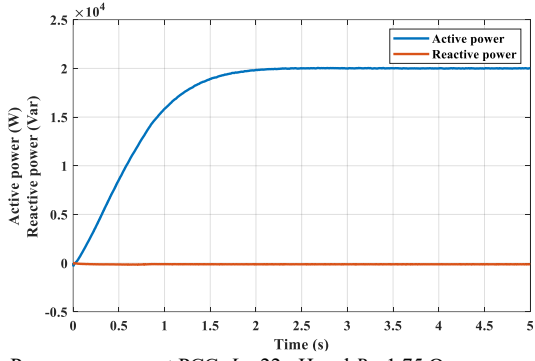

Fig. 14. Power responses at PCC: $L_g$=22mH and $R_g$=1.75 Ω.

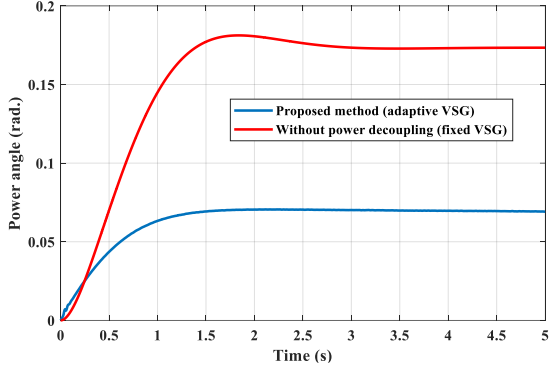

Fig. 15. Power angle: $L_g$=22mH and $R_g$=1.75 Ω.

Table V
Power angels under different grid impedance conditions

| Grid impedance parameters | | With Coupling (rad.) | Proposed Method (rad.) |
|---|---|---|---|
| R/X=1 | $R_g$=0.75 Ω, $L_g$=2.3 mH | $δ$=0.12 | $δ$=0.08 |
| | $R_g$=1.75 Ω, $L_g$=5.5 mH | $δ$=0.28 | $δ$=0.17 |
| R/X=2.5 | $R_g$=0.75 Ω, $L_g$=0.95 mH | $δ$=0.05 | $δ$=0.025 |
| | $R_g$=1.75 Ω, $L_g$=2.2 mH | $δ$=0.17 | $δ$=0.07 |

Table III
The percentage increase of power transferred from the inverter

| Grid impedance parameters | | Increase in active power delivery to the grid |
|---|---|---|
| R/X=1 | $R_g$=0.75 Ω, $L_g$=2.3 mH | 0.8 % |
| | $R_g$=1.75 Ω, $L_g$=5.5 mH | 6 % |
| R/X=2.5 | $R_g$=0.75 Ω, $L_g$=0.95 mH | 1.1 % |
| | $R_g$=1.75 Ω, $L_g$=2.2 mH | 6.7 % |

### C. Step change in local load

The proposed method can also decouple power in the presence of a local load at PCC. Figs. 16 and 17 show the inverter's power responses when a local load with 5 kW is applied at t=5 sec. for both cases, i.e., R/X=1 and R/X=2.5, respectively.

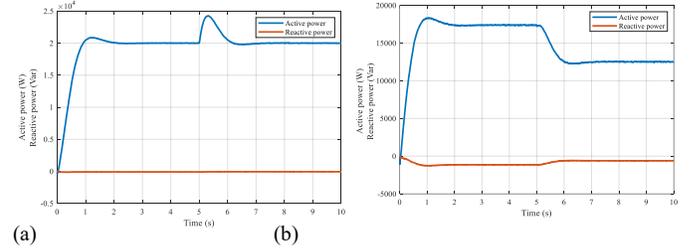

Fig. 16. Power responses: $L_g$=4mH and $R_g$=1.25 Ω (a) at PCC (b) at grid.

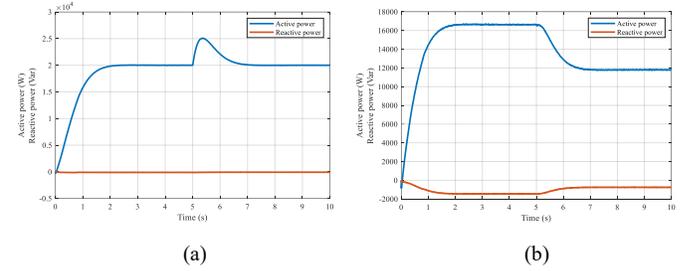

Fig. 17. Power responses: $L_g$=22mH and $R_g$=1.75 Ω (a) at PCC (b) at grid.

Table VII
Comparison of different power decoupling methods

| Decoupling method | | Considered aspects | | | | Assessments indicators | | | | | |
|---|---|---|---|---|---|---|---|---|---|---|---|
| | | Change in R/X | Change in $Z_g$ | $δ$ | VSG control | Clear physical meaning | Esay implementation | Decoupling performance | Require prior knowledge of system | Need known grid impedance | VSG adaptive feature |
| Virtual impedance | [6] | × | × | × | ✓ | High | High | Low | No | Yes | No |
| Hybrid | [17] | ✓ | × | × | ✓ | High | High | Medium | Yes | Yes | Yes |
| Feedforward | [10] | × | × | ✓ | ✓ | Medium | Medium | High | Yes | Yes | No |
| | [11] | × | ✓ | ✓ | ✓ | Medium | High | High | Yes | Yes | No |
| Proposed method | | ✓ | ✓ | ✓ | ✓ | High | High | High | No | Yes | Yes |

Note: the symbol ✓ denotes the positive property, and × denotes the negative property.

## D. Step change in in active power reference

Figs. 18 illustrates the inverter's power response as the reference value of active power are changes. At t=3 sec., the reference active power value changes from 10 kW to 20 kW, while t=6 sec., it drops back to the initial reference value (10 kW). This result demonstrates that the proposed method easily handles the change in the reference value of the active power.

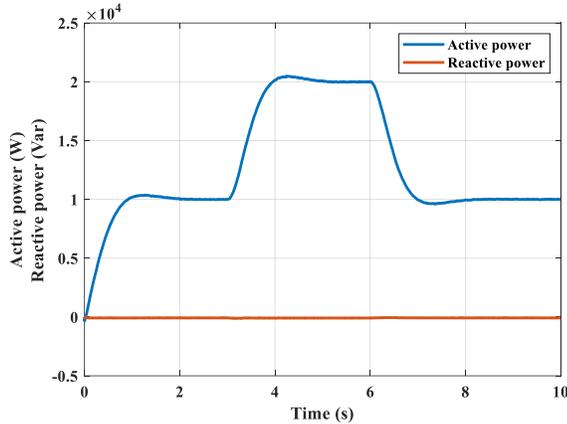

Fig. 18. Power responses at PCC when step change is applied in the active power reference value: *Lg=4mH and Rg=1.25 Ω*.

## VI. Conclusion

This paper proposes a new vision for power decoupling in VSG-based grid-forming inverters. This vision uses a fuzzy logic technique to adjust power control loop parameters: *kq, J* and *D*. The results showed that the dynamic adaptation of these parameters leads to eliminating both static and dynamic power coupling and increases the performance of the grid-forming inverter and thus contributes to the stability of the system and provide active power delivery capability to the grid. Moreover, due to its adaptive facility, this inverter can support the grid with adaptive synchronous generator characteristics under different operating conditions. Another unique feature of the proposed method is that it does not require prior knowledge of the system model (model-free based method). The proposed method was compared with state of the art through several criteria, where it can be said that it represents a promising adaptive method-based power control for grid-forming inverter.